[1999/12/01 v1.4c Il Nuovo Cimento]
\documentclass{cimento}

\usepackage{graphicx}% Include figure files
\usepackage{bm}% bold math
\usepackage{epsfig}

\begin{document}

%\headnote{Wealth 1.0}
%%\preprint{S N Wealth, Version: 1.0}
%\offprints{}

\title{
Exchanges in complex networks: income and wealth distributions
}

\author{
T. Di Matteo\from{ins:x}\from{ins:y}, T. Aste\from{ins:x} and S. T. Hyde\from{ins:x}
}

\instlist{
\inst{ins:x} Applied Mathematics, Research School of Physical Sciences, Australian National University, 0200 Canberra, Australia. tdm110@rsphysse.anu.edu.au; tas110@rsphysse.anu.edu.au; sth110@rsphysse.anu.edu.au.
\inst{ins:y} INFM-Dipartimento di Fisica ``E. R. Caianiello'', via S. Allende, 84081 Baronissi (SA) Italy. 
}

\PACSes{\PACSit{89.65.Gh }{ Economics; econophysics, financial markets, business and management }
\PACSit{ 89.90.+n }{ Other topics in areas of applied and interdisciplinary physics }
}

\maketitle

\begin{abstract}
We investigate the wealth evolution in a system of agents that exchange wealth through a disordered network in presence of an additive stochastic Gaussian noise. 
We show that the resulting wealth distribution is shaped by the degree distribution of the underlying network and in particular we verify that scale free networks generate distributions with power-law tails in the high-income region.
Numerical simulations of wealth exchanges performed on two different kind of networks show the inner relation between the wealth distribution and the network properties and confirm the agreement with a self-consistent solution.
We show that empirical data for the income distribution in Australia are qualitatively well described by our theoretical predictions.
\end{abstract}

\section{Introduction}

\begin{figure}
\begin{center}
\begin{tabular}{cc}
\mbox{\epsfig{file=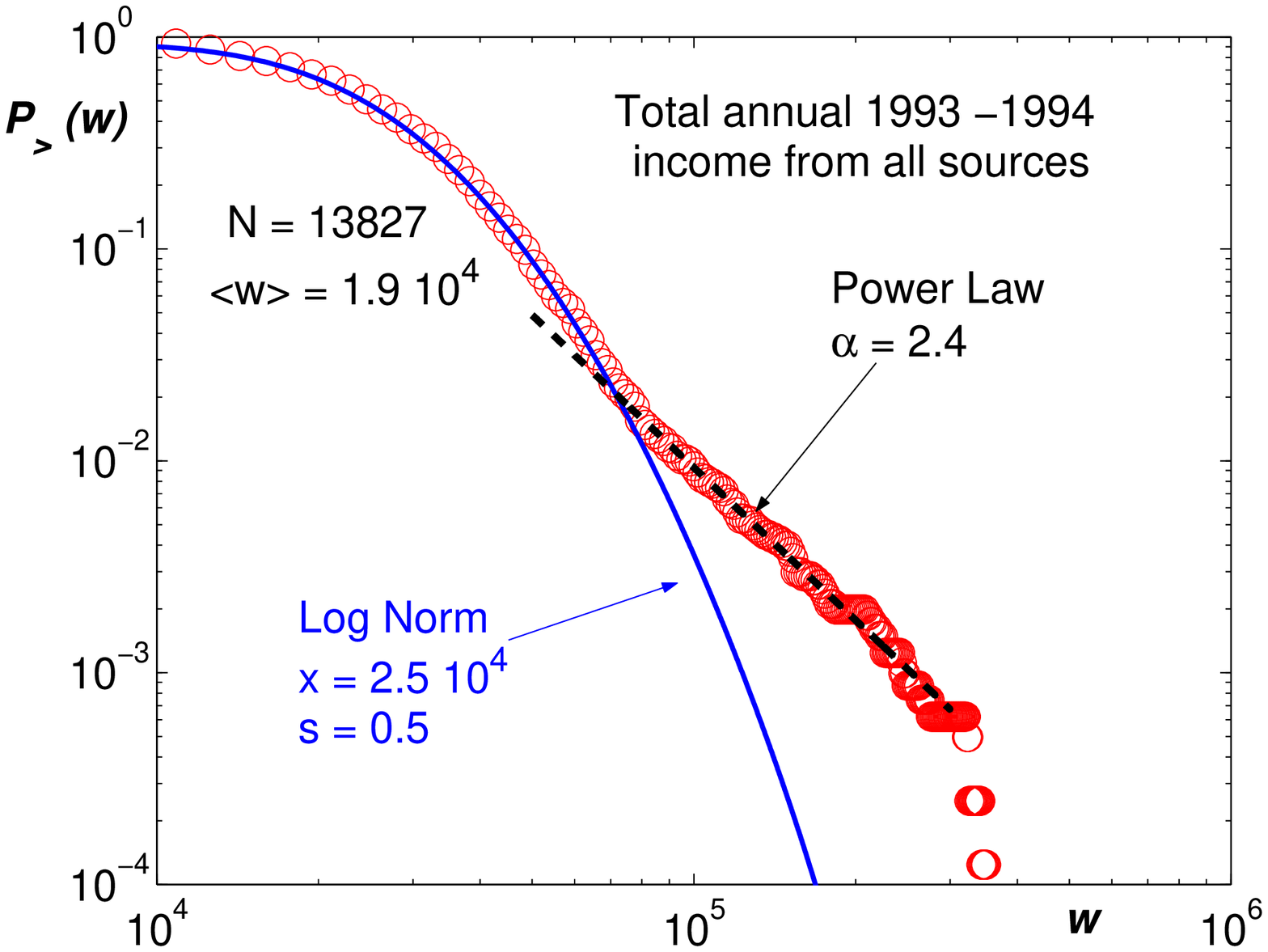,width=6.cm,angle=0}}
&\mbox{\epsfig{file=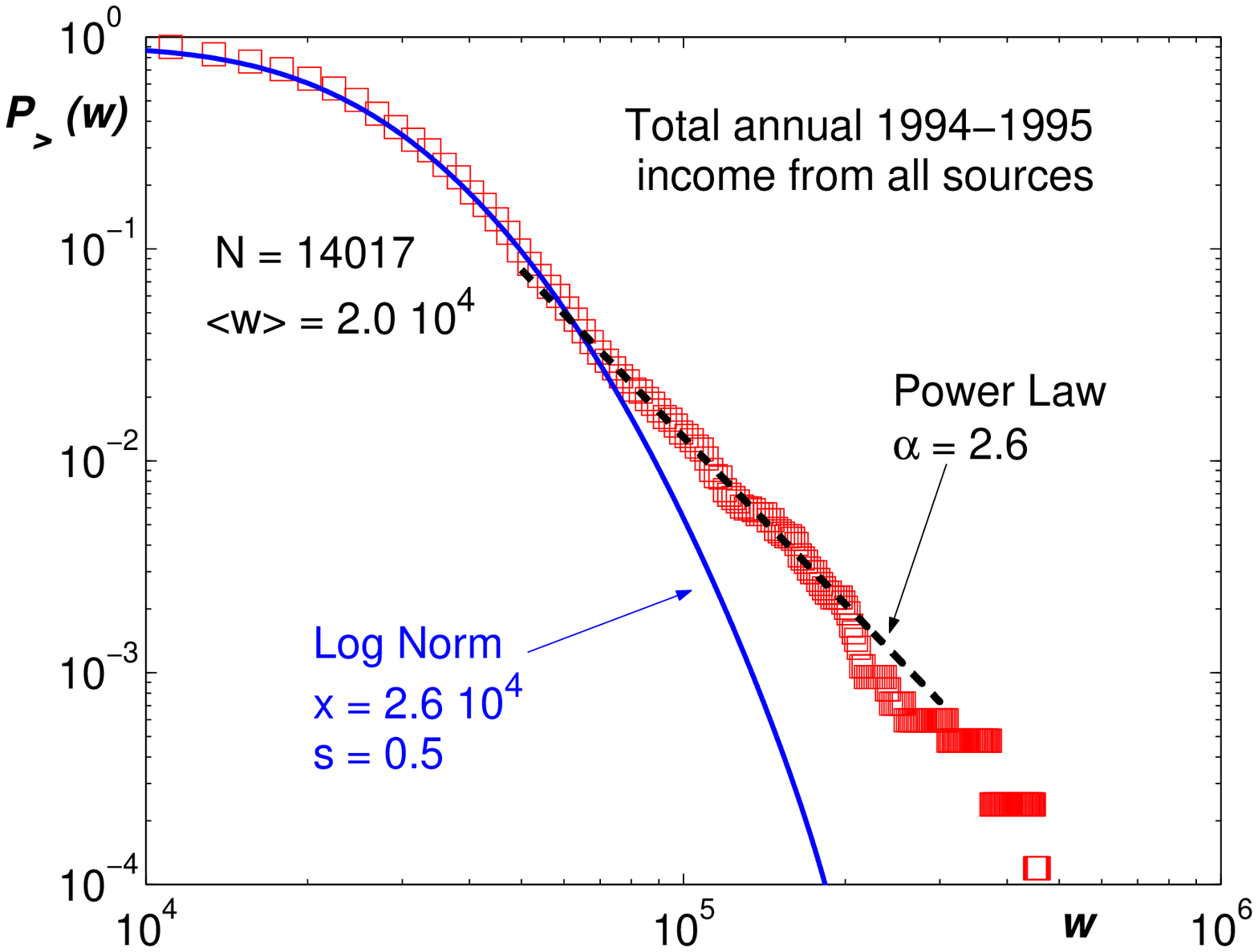,width=6.cm,angle=0}}\\
\mbox{\epsfig{file=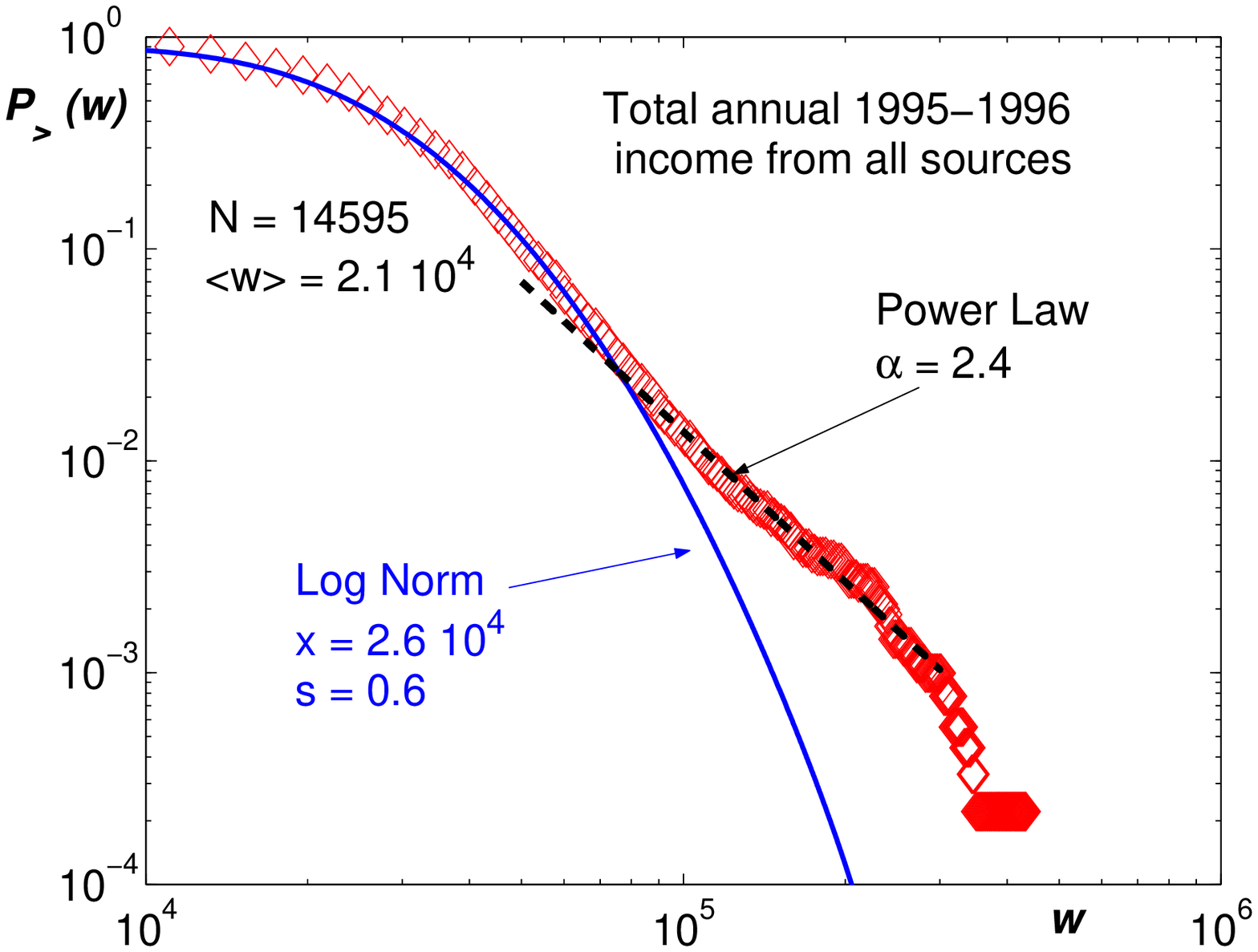,width=6.cm,angle=0}}
&\mbox{\epsfig{file=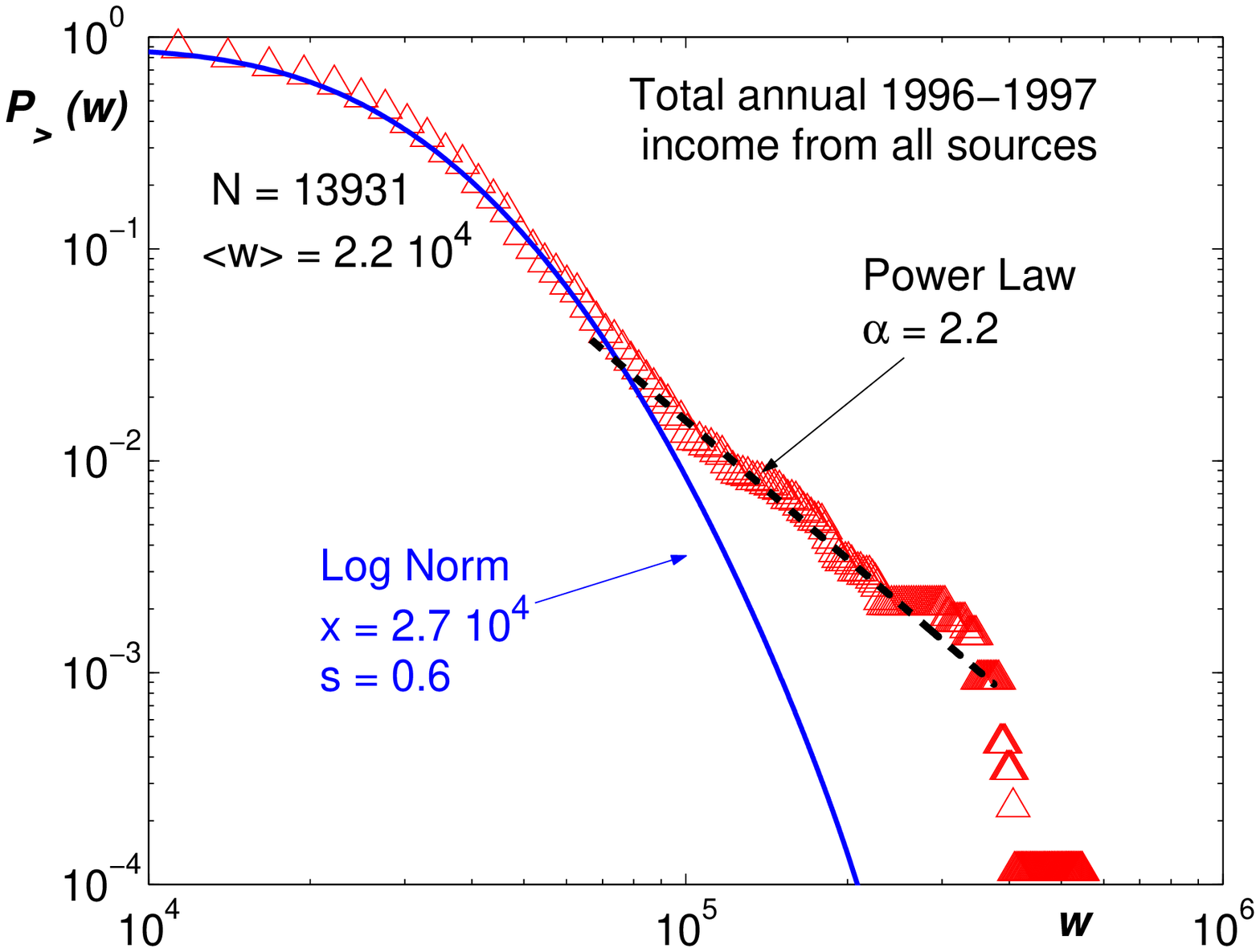,width=6.cm,angle=0}} 
\end{tabular}
\end{center}
\caption{ 
Complementary cumulative distributions for the Total annual income from all sources in Australia in the years $1993-1997$. 
}
\label{f.income}
\end{figure}

Empirically the literature reports several behaviors for the income and wealth distributions in different countries.
A century ago, the Italian social economist Pareto suggested a power-law~\cite{Stanley2003} distribution in the high-income range, namely, in terms of cumulative distribution: $P_> (w) \propto w^{-\alpha}$, with $\alpha$ being the Pareto index~\cite{Pareto1897}. 
On the other hand Montroll~\cite{Montroll1974} suggested a lognormal distribution with power law tail for the USA personal income. 
More recently, wealth and income distributions in the USA and in the United Kingdom have been described by an exponential distribution with power law high-end tails~\cite{Dragulescu2001}. 
Whereas, the Japanese personal income distribution appears to follow lognormal distributions also with power law tails~\cite{Souma2001,Fujiwara2003}. 
In some recent papers Zipfs law has also been proposed~\cite{Aoyama2000}.
In this paper we add to the above empirical investigations an analysis of the income distribution in Australia (Figure~\ref{f.income}).

>From the theoretical side, it has been shown that pure multiplicative stochastic (MSP) processes can explain the lognormal income distribution but they fail to explain the power law tails~\cite{Gibrat1932}.
Power law tails can be obtained extending MSP processes by including -for instance- additive noise and boundary constraints ~\cite{Levy1996,Sornette1998,Huang2002,Richmond2001}. 
These models explain well the emergence of power law distributions, but they are incomplete, neglecting interactions between agents. 
Hence, MSPs with interacting agents connected through a network have been developed~\cite{Stanley1998,Biham1998,Bouchaud2000,Solomon2002}. 
These models retrieve power law tails with exponents $\alpha$ which are related to the network properties.

In this paper, we show that distributions with power law tails can emerge also from additive stochastic processes with interacting agents.
In this case, we show that the network of connections among agents plays a crucial role.
Indeed, the resulting wealth distribution is shaped directly by the degree distribution of the network.
The original purpose of the present work was not to construct any realistic model for the wealth distribution.
Our aim was simply to demonstrate the possibility to obtain `fat' tails also without the use of multiplicative stochastic processes.
Rather surprising we find out that the results from such an additive process are in good qualitative agreement with the empirical data for the income distribution in Australia. 

%\vspace*{3cm}

\section{Income distribution in Australia}

Let us briefly start with the empirical analysis of the data for the incomes in Australia.
We analyze data from the Australian Bureau of Statistics: ``Survey of Income and Housing Costs Confidentialised Unit Record Files''. 
In Figure~\ref{f.income} we report the complementary cumulative distributions ($P_>(w) = 1- \int_{-\infty}^{w} p(\xi) d \xi$) for the Total annual income from all sources in the years $1993-94$, $1994-95$, $1995-96$, $1996-97$. 
These data are compared with two possible trends in two different regions: lognormal at low and medium incomes and power law at high incomes.
As one can see the large income region is rather well described with power law -like tails: $P_>(w) \propto w^{-\alpha}$ with exponents $\alpha$ respectively equal to $2.4$, $2.6$, $2.4$, $2.2$.
Whereas the small incomes region is in better agreement with a lognormal distribution: $P(w)=1/(w s \sqrt{2 \pi})\exp[-\log^2(w/x)/(2 s^2)]$ (with the values for $s$ and $x$ reported in the figures).
Let us now introduce the theoretical framework and show how these behaviors can be accounted by using an additive interacting stochastic process.

%\vspace*{3cm}
\begin{figure}
\begin{center}
\mbox{\epsfig{file=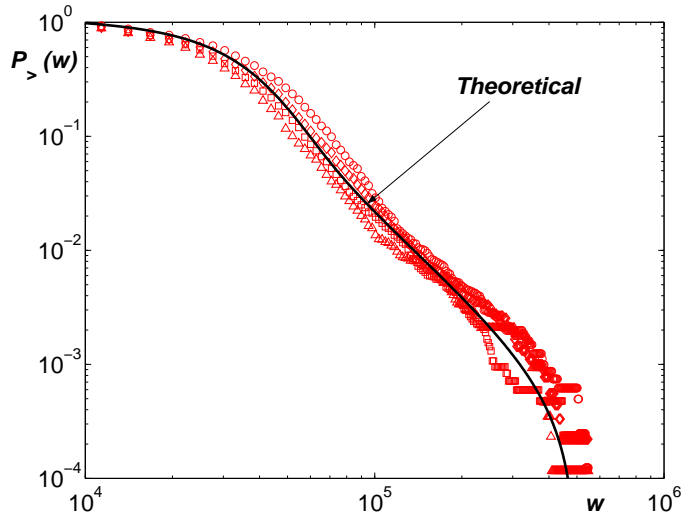,width=9.cm,angle=0}} 
\end{center}
\caption{ 
Comparison between the empirical data and the theoretical (complementary cumulative) distribution associated with a scale-free network.
}
\label{f.comp}
\end{figure}

\section{Wealth distribution from interacting additive stochastic processes}

Consider $N$ agents which interact through a social network and suppose that at the time $t$ a given agent $l$ has a wealth $w_l(t)$. 
Within the same framework of other models proposed in the literature ~\cite{Biham1998,Bouchaud2000,Solomon2002}, let us first introduce a rather general expression for the wealth evolution:
\begin{equation} 
\label{W}
w_l(t+1)- w_l(t) = A_l(t)+B_l(t) w_l(t) + \sum_{j (\neq l) } w_j(t) Q_{j \to l}(t) - \sum_{j (\neq l) } w_l(t) Q_{l \to j}(t) \;\;\;\;,
\end{equation}
where the coefficient $A_l(t)$ is an additive noise and the factor $B_l(t)$ is a multiplicative noise. 
These are stochastic variables which reflect market and social fluctuations.
In addiction with these stochastic terms Equation~\ref{W} describes the exchanges between agents through a network: agent $l$ receives a faction $Q_{j \to l}(t)$ of the wealth of agent $j$ and gives a fraction $Q_{l\to j'}(t)$ of its wealth to agent $j'$.
The MSP model mentioned above takes into account only the multiplicative term $B_l(t)$; their extensions introduce also the additive noise $A_l(t)$ and the interactions $Q_{j \to l}(t)$.
Differently, in this paper we neglect the multiplicative term and take into account only the additive noise and the interactions.
In particular we assume that: 
i) there are no stochastic multiplicative terms ($B_l(t)=0$); 
ii) the additive term $A_l(t)$ is a Gaussian noise with average zero and variance $\sigma_0^2$; 
iii) each agent distributes a portion $q_0$ of its wealth equally among the other agents which are in contact with it through the social network. 
This last assumption implies:
\begin{equation}\label{EqSh}
Q_{j\to l}(t) = 
\left\{ 
\begin{array}{ll}
 \frac{\textstyle q_0}{\textstyle z_j} & \mbox{if $l \in \mathcal{I}_j$; } \\
0 & \mbox{elsewhere.}
\end{array}
\right.
\end{equation}
where $z_j$ is the number of agents in contact with agent $j$ and $\mathcal{I}_j$ represents the set of the agents which exchange with agent $j$. 
Equation~\ref{W} becomes
\begin{equation}\label{W1}
w_l(t+1) = A_l(t)+ ( 1- q_0) w_l(t) + \sum_{j \in \mathcal{I}_l} \frac{q_0}{z_j} w_j(t) \;\;\;\;.
\end{equation}
Note that in our case $\left< A_l(t) \right>_t = 0$ and Equation~\ref{W1} describes a system which conserves in average the total wealth.

The probability $P_{t+1}(x,l)dx$ that the agent $l$ at the time $t+1$ has a wealth between $x$ and $x+dx$ is related to the probabilities to have a set $\{Q_{j\to l}(t)\}$ of exchange coefficients and a set of additive coefficient $\{A_l(t)\}$ such that a given distribution of wealth $\{w_j(t)\}$ at the time $t$ yields, through Equation~\ref{W1}, to the wealth $x$ for the agent $l$ at time $t+1$.
This is:
\begin{equation}\label{Pw-1}
P_{t+1}(x,l) = 
\int_{-\infty}^\infty \! da \,
\Lambda_t(a,l) 
\int_{-\infty }^\infty \!\! dw_1 \cdots \!
\int_{-\infty }^\infty \!\! dw_N 
P_{t}(w_j,j)
\delta \big(x - a - (1- q_0) w_l - \sum_{j'} \frac{q_0}{z_{j'}} w_{j'} \big) ,
\end{equation}
where $\delta(x)$ is the Dirac delta function and $\Lambda_t(a,l)$ is the probability density to have at time $t$ on site $l$ an additive coefficient $A_l(t)=a$.

The Fourier transform of Equation~\ref{Pw-1} reads:
\begin{equation}\label{Pw-2}
\hat P_{t+1}(\varphi,l) 
= 
\frac{ e^{ -\frac{\sigma_0^2 \varphi^2}{2} }}{\sqrt{2\pi}} 
\hat P_t((1-q_0)\varphi,l) 
\prod_{j \in \mathcal{I}_l}
\hat P_t(\frac{q_0}{z_j} \varphi,j) 
\;\;.
\end{equation}

By definition the cumulants of the wealth probability distribution are given by the expression:
\begin{equation}\label{C-1}
k^{(\nu)}_l(t) = (-i)^{\nu}\frac{d^{\nu}}{d\varphi^{\nu}} \ln \hat P_t(\varphi,l) \Big|_{\varphi=0} \;\;,
\end{equation}
where the first cumulant $k^{(1)}_l(t)$ is the expectation value of the stochastic variable $w_l$ at the time $t$ ($\left< w_l(t) \right> $) and the second moment $k^{(2)}_l(t)$ is its variance ($\sigma_l^2(t)$).

By taking the logarithm of Equation~\ref{Pw-2} and applying Equation~\ref{C-1} we get:
\begin{equation}\label{C-2}
k^{(\nu)}_l(t+1) = c^{(\nu)} 
+ (1-q_0)^\nu k^{(\nu)}_l(t) 
+ \sum_{j \in \mathcal{I}_l} \left(\frac{q_0}{z_j}\right)^\nu k^{(\nu)}_j(t) \;\;\;,
\end{equation}
with $ c^{(2)}=\sigma_0^2$ and $c^{(\nu)}=0$ for any $\nu \not=2$. 
This equation describes the propagation of the cumulants of the wealth distribution. 
A consequence of this equation is that the only moments which asymptotically can be different from zero are the first (the mean) and the second (the variance).
With this last being directly proportional to $\sigma_0^2$. 

We now seek for stationary solutions of Equation~\ref{C-2}, i.e. situations in which at infinitely large times, the cumulants do not change in time: $k^{(\nu)}_l(t)= \bar k^{(\nu)}_l$.

%\vspace*{3cm}
\begin{figure}
\begin{center}
\begin{tabular}{cc}
\mbox{\epsfig{file=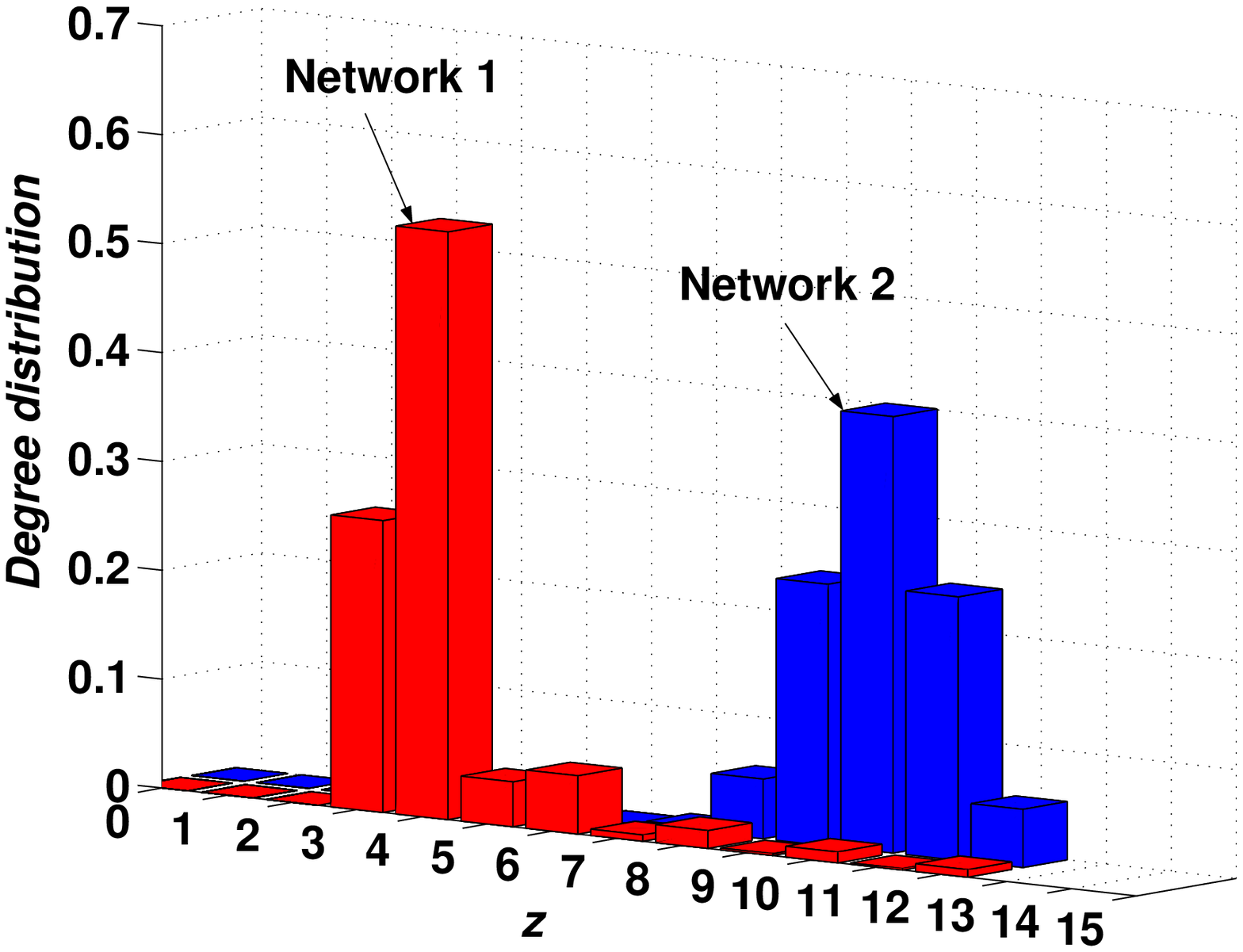,width=6.cm,angle=0}} 
&
\mbox{\epsfig{file=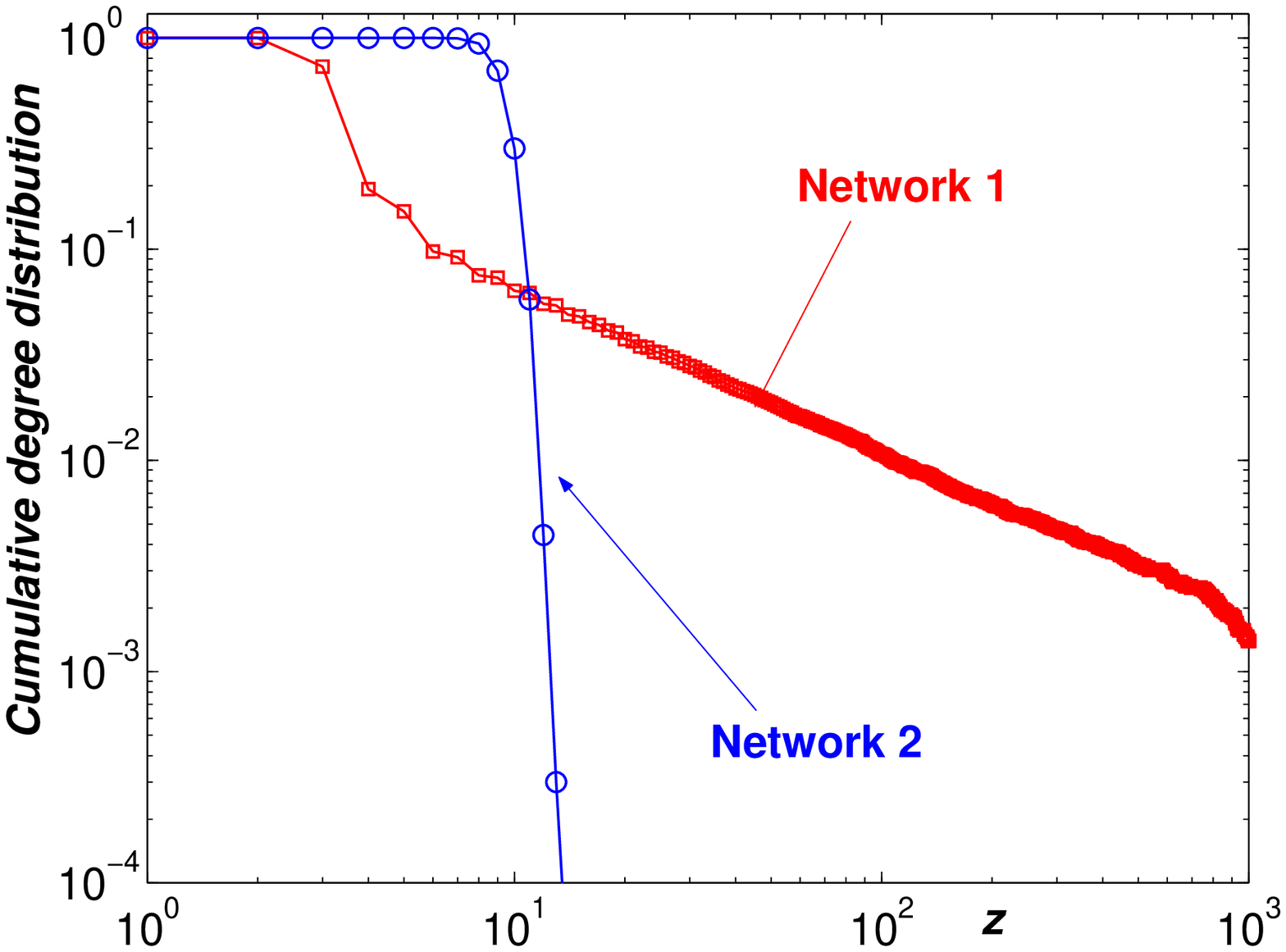,width=6.cm,angle=0}} 
\end{tabular}
\end{center}
\caption{ 
The two degree distributions for the two networks used in the simulations. (Left: degree distributions, Right: complementary cumulative degree distributions).
}
\label{f.degree}
\end{figure}

\subsection{Self-Consistent solution: crystal}

Let us first consider an `ideal' social network where every agent is connected with an equal number of other agents ($z_j = \bar z$) in a perfectly ordered `crystalline' structure. 
In this case, each agent is equivalent to each other and the asymptotic wealth distribution must be the same for every one (i.e. $k^{(\nu)}_j$ independent on $j$).
>From Equation~\ref{C-2} it follows that the expectation value for the wealth on each site is a constant and it is equal to the average wealth at $t=0$:
\begin{equation}\label{Av-Cry}
\left< w_l \right> = \bar k_l^{(1)} = \frac{1}{N} \sum_j w_j(0) \;\;\;.
\end{equation}
Its variance is
\begin{equation}\label{C-cry}
\sigma^2_l = \bar k_l^{(2)} = \frac{\sigma_0^2}{1-\frac{q_0^{2}}{\bar z} - (1- q_0)^2} \;\;\;.
\end{equation}
Whereas all the other moments $\bar k^{(\nu)}$ are equal to zero for $\nu \geq 3$.
%\vspace*{3cm}
\begin{figure}
\begin{center}
\begin{tabular}{cc}
\mbox{\epsfig{file=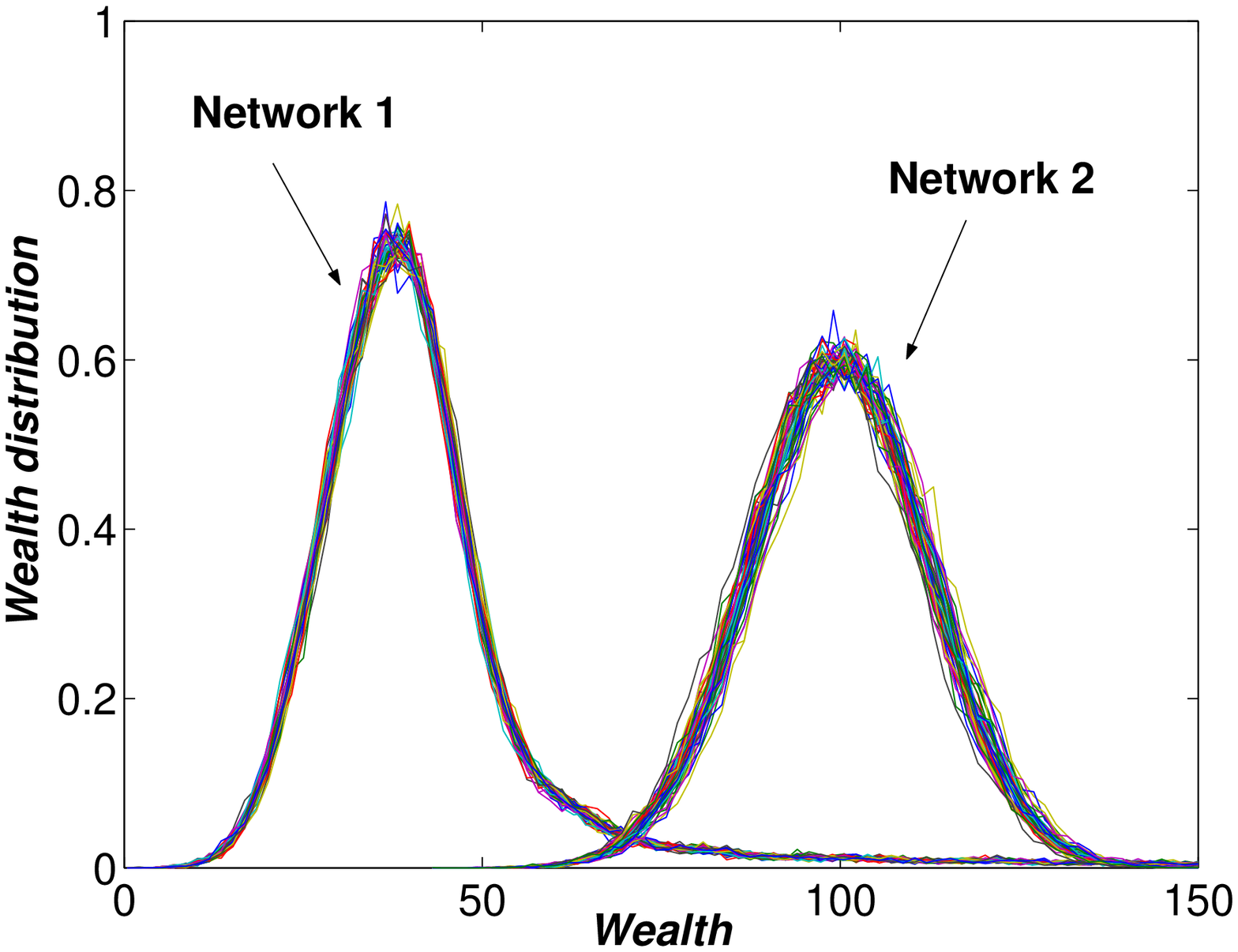,width=6.cm,angle=0}} 
&
\mbox{\epsfig{file=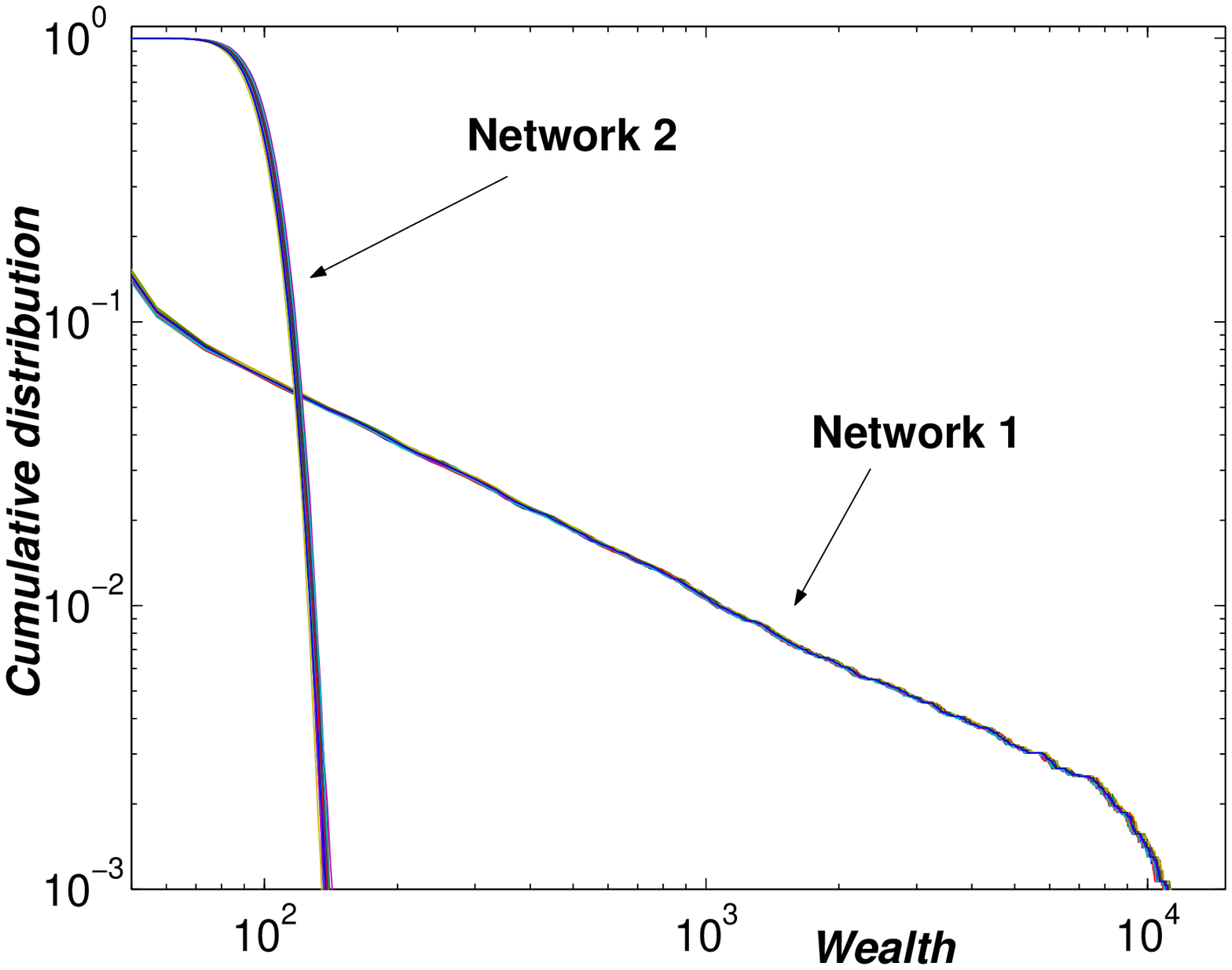,width=6.cm,angle=0}} 
\end{tabular}
\end{center}
\caption{ 
The wealth distributions (left) and their complementary cumulative distributions (right)  resulting from 50 simulations performed on two the different networks (1 and 2). }
\label{f.wealth}
\end{figure}

\subsection{Self-Consistent solution: general case}

We now consider the more general case of a non-regular network.
>From Equation~\ref{C-2} it follows that a self-consistent solution for the average stationary wealth on each node of the network is:
\begin{equation}\label{C-4}
\left< w_l \right> = \bar k_l^{(1)} = \frac{z_l}{\bar z} m \;\;\;,
\end{equation}
with $m$ the average wealth on the ensemble of agents ($m = \frac{1}{N} \sum_j w_j(t)$) and $\bar z$ the average network connectivity ($\bar z = \frac{1}{N} \sum_j z_j$).
Therefore the expectation value for the wealth of a given agent results proportional to its number of connections in the social network.
On the other hand, we mentioned above that the only other moment which can be different from zero is the second.
Therefore, we expect that the probability to find a given wealth on a given agent is a Gaussian distribution with average $\frac{z_l}{\bar z} m$ and finite variance.

\subsection{Wealth distribution -analytical}

The wealth probability distribution in the ensemble of agents is given by the sum of the distributions for each agent divided by the total number of agents.
We have seen above that each agent has a wealth within a Gaussian distribution with average directly proportional to its connectivity (Equation~\ref{C-4}) and finite variance. 
The resulting wealth distribution for the ensemble of agent is therefore a weighted sum of Gaussian distributions with averages proportional to the network connectivity and weights given by the degree distribution. 
This overall distribution is shaped by the underlying distribution of the connectivity between agents (the degree distribution).
It has been observed that in many social systems the degree distribution typically follows a power law behavior in the region of large number of connections~\cite{Newman2003}.
This power law behavior in the social network connectivity will be therefore reflected in the wealth distribution which will assume a power law tail terminated by an exponential cutoff (for a finite system).
This behavior is qualitatively in agreement with what observed empirically.
A comparison between the empirical data and the distribution resulting by summing a set of $3000$ Gaussian distributions with averages proportional to the connectivity $x = \frac{z}{\bar z} x_0$ (with $x_0=30000$, $\bar z =1.28$), equal variances $s = 18000$ and power law degree distribution $p(z)=p_0 z^{-a}$  (with $a=3.2$, $z_{min}=1$, $z_{max}=3000$ ) is shown in Figure~\ref{f.comp}.
As one can see the qualitative agreement is quite satisfactory.

\subsection{Wealth distribution - numerical simulation}

We generated large networks, with $N=30000$ agents, by iteratively performing switching of neighbors in a triangulation embedded in a manifold with genus $g=10000$.
We introduce an `energy' $E = \sum_j (z_j-\bar z)^2$ and we perform a Glauber-Kawasaki type of dynamics. 
This procedure is an extension to $g \not= 1$ and negative `temperatures' of the method presented in ~\cite{Aste00}. 
Two different networks were generated by performing 600000 switches from a disordered start respectively at inverse `temperatures' $\beta = - 0.5$ (network 1) and $\beta = + 0.5$ (network 2). 
At positive temperature a rather homogeneous network emerges with degree distribution centered around the average ($\bar z = 6 + 12 (g-1)/N$) and with exponentially fast decreasing tails (Figure~\ref{f.degree}, network 2). 
On the other hand, negative temperatures favour the formation of inhomogeneous-scale free networks, with power law tails in the degree distribution (Figure~\ref{f.degree}, network 1).

\begin{figure}
\begin{center}
\begin{tabular}{cc}
\mbox{\epsfig{file=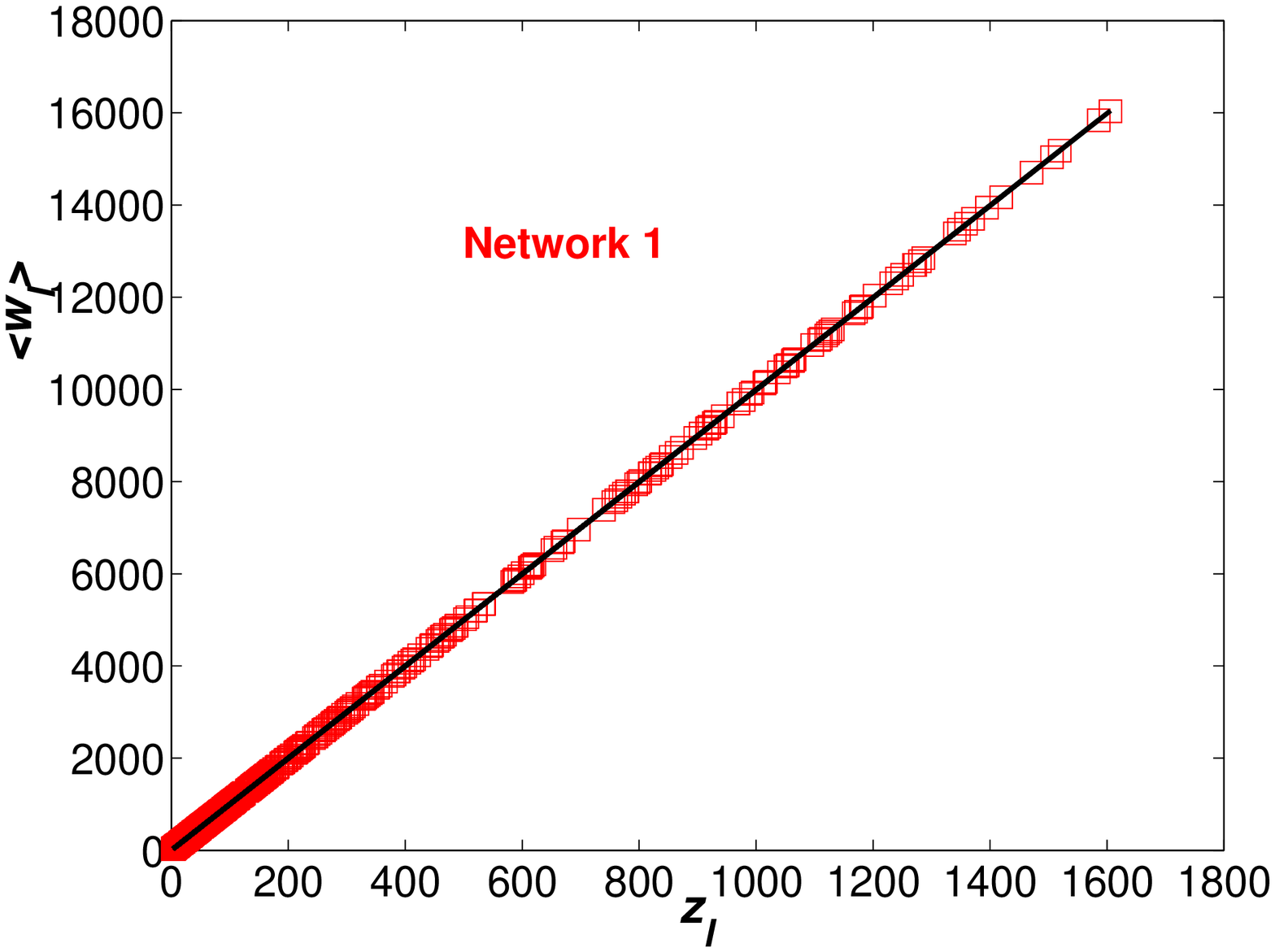,width=6.cm,angle=0}} 
&
\mbox{\epsfig{file=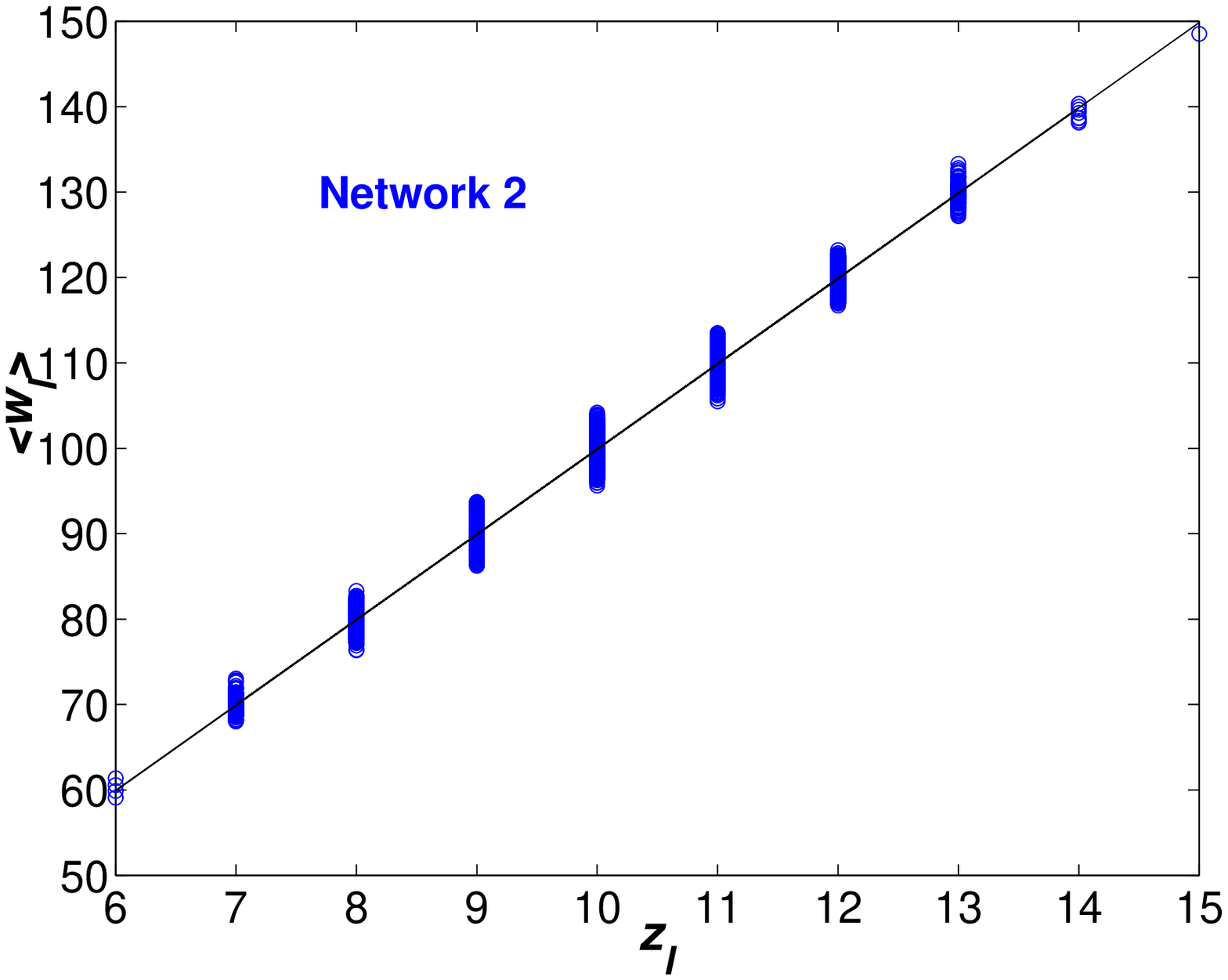,width=6.cm,angle=0}} 
\end{tabular}
\end{center}
\caption{ 
Comparison between the average values for the wealth of a given agent $l$ (symbols) calculated from 50 simulations on the two networks (left: network 1; right: network 2) with the theoretical predictions from Equation~\ref{C-4} (lines).
}
\label{f.meanfield}
\end{figure}

Once the networks are generated, we associate to each agent an equal initial wealth of $m$ (arbitrary) units.
We set $m=100$, $q_0 = 0.1$, $\sigma = 0.05/m$ and we run the simulation by updating at each time-step all the agent's wealth by using Equation~\ref{W1} up to a maximum time $T$.
We verify that a steady state distribution is achieved after about 100 time-steps and therefore we set $T=1000$.
The resulting wealth distributions for 50 simulations over fixed underlying networks are reported in Figure~\ref{f.wealth}.
We see that the network properties have a dramatic effect on the overall behavior of the wealth distribution.
We observe an exponentially fast decay in the homogeneous network, whereas we obtain a power law `fat' tail in the scale-free one.
We verify that, in agreement with Equation~\ref{C-4}, the expectation velue for the wealth on each agent is proportional to its connectivity in the social network.
Figure~\ref{f.meanfield} reports the average (over the 50 simulation) wealth on each agent v.s. its connectivty.
The theoretical prediction (Equation~\ref{C-4}) is also reported showing a remarkable agreement for both the networks.
The small spreading of the data (more evident for network 2) indicates that non-local effects might also have some relevance.

%\vspace*{3cm}

\section{Conclusion}

We have shown that a mechanism of wealth exchange with additive Gaussian noise can produce distributions with power-law tails when the network which connects the agents is of a scale-free type.
Although the original purpose of this work was not to produce a realistic model for the wealth evolution, we find a good qualitative agreement between the empirical data and the theoretical prediction. 
More realistic models will be proposed in future works by introducing also multiplicative stochastic terms and a dynamical evolution in the network connectivity.

\acknowledgments
T. Di Matteo wishes to thank the Research School of Social Sciences, ANU, for providing the ABS Data.
This work was partially founded by the ARC Discovery Project: DP0344004.
We acknowledge the STAC Supercomputer Time Grant at APAC National Facility.

%\newpage


\begin{thebibliography}{0}

\bibitem{Stanley2003}
\BY{X. Gabaix, P. Gopikrishnan, V. Plerou and H. E. Stanley} 
%A theory of power-law distributions in financial market fluctuations, 
\IN{Nature}{423}{2003}{267-270}.

\bibitem{Pareto1897}
\BY{V. Pareto}
\TITLE{ {Cours d' \`Economique Politique} }
(Macmillan, London) 1897.

\bibitem{Montroll1974}
\BY{E. W. Montroll and W. W. Badger} 
\TITLE{Introduction to Quantitative Aspects of Social Phenomena}
(Gordon and Breach, New York) 1974.

\bibitem{Dragulescu2001}
\BY{A. Dragulescu and V. M. Yakovenko} 
%Evidence for the exponential distribution of income in the USA, 
\IN{Eur. Phys. J.}{B 20}{2001}{585-589} 
and 
\IN{Physica A}{299}{2001}{213-221}.

\bibitem{Souma2001}
\BY{W. Souma} 
%Universal structure of the personal income distribution, 
\IN{Fractals}{9}{2001}{463-470}.

\bibitem{Fujiwara2003}
\BY{Y. Fujiwara, W. Souma, H. Aoyama, T. Kaizoji, M. Aoki} 
%Growth and fluctuations of personal income, 
\IN{Physica A}{321}{2003}{598-604}.

\bibitem{Aoyama2000}
\BY{H. Aoyama, W. Souma, Y. Nagahara, M. P. Okazaki, H. Takayasu, N. Takayasu} %Paretoís law for income of individuals 
%and debt of bankrupt companies, 
\IN{Fractals}{8}{2000}{293-300}.

\bibitem{Gibrat1932}
\BY{R. Gibrat} 
\TITLE{Les In\'egalit\'es \'Economiques}
(Sirey, Paris, 1932).

\bibitem{Levy1996} 
\BY{M. Levy, S. Solomon} 
%Power laws are logarithmic Boltzman laws, 
\IN{Int. J. Mod. Phys.}{C7}{1996}{595-601}.

\bibitem{Sornette1998}
\BY{D. Sornette}
%Multiplicative processes and power laws, 
\IN{Phys. Rev. E}{57}{1998}{4811-4813}.

\bibitem{Huang2002}
\BY{Zhi-Feng Huang, S. Solomon} 
\IN{Physica A}{294}{2001}{503-513}. 

\bibitem{Richmond2001}
\BY{S. Solomon, P. Richmond}
%Power law of wealth, market order volumes and market returns, 
\IN{Physica A}{299}{2001}{188-197}.

\bibitem{Stanley1998}
\BY{L. A. N. Amaral, S. V. Buldyrev, S. Havlin, M. A. Salinger, and H. E. Stanley} %Power Law Scaling for a System of Interacting Units with Complex Internal Structure, 
\IN{Phys. Rev. Lett.}{80}{1998}{1385-1388}.

\bibitem{Biham1998} 
\BY{O. Biham, O. Malcai, M. Levy and S. Solomon}
%Generic emergence of power law distributions and Levy-stable intermittent fluctuations in discrete logistic systems, 
\IN{Phys. Rev. E}{58}{1998}{1352-1358}.

\bibitem{Bouchaud2000}
\BY{J.-P. Bouchaud, M. M\'ezard}
 \IN{Physica A}{282}{2000}{536-545}.

\bibitem{Solomon2002}
\BY{S. Solomon and P. Richmond}
\IN{Eur. Phys. J. B}{27}{2002}{257-261}.

\bibitem{Newman2003}
\BY{M. E. J. Newman}
\IN{SIAM Review}{45}{2003}{167-256}.

\bibitem{Aste00}
\BY{T. Aste and D. Sherrington}
\IN{ J. Phys. A: Math. Gen.}{ 32}{1999}{7049-7056}.



\end{thebibliography}
\end{document}